\documentclass[useAMS,usenatbib,letters]{mn2e}
\usepackage{amsmath}
\usepackage{natbib}
\usepackage{graphicx}
\usepackage{color}
\usepackage{txfonts}
\usepackage{subfigure}
\usepackage{nicefrac}
\usepackage{xspace}
\usepackage{url}
\usepackage{amssymb}

\newcommand{\reb}{\texttt{REBOUND}\xspace }
\newcommand{\rev}[1]{#1}

\title{Period ratios in multi-planetary systems discovered by Kepler are consistent with planet migration}
\author[Hanno Rein]{Hanno Rein\\
\\
Institute for Advanced Study, 1 Einstein Drive, Princeton, NJ 08540, e-mail: \url{rein@ias.edu}}
\date{Received: 26 July 2012. Accepted: 16 August 2012.}

\begin{document}
\maketitle
\begin{abstract}
The Kepler planet candidates are an interesting testbed for planet formation scenarios. 
We present results from $N$-body simulations of multi-planetary systems that resemble those observed by Kepler.
We add both smooth (Type I/II) and stochastic migration forces. 
The observed period ratio distribution is inconsistent with either of those two scenarios on its own.

However, applying both stochastic and smooth migration forces to the planets simultaneously results in a period ratio distribution that is similar to the observed one.
This is a natural \rev{scenario} if planets form in a turbulent proto-planetary disk where these forces are always present. 
We show how the observed period ratio and eccentricity distribution can constrain the relative strength of these forces, a parameter which has been notoriously hard to predict for decades.

We make the source code \rev{of our simulations} and the initial conditions freely available to enable the community to expand this study and include effect other than planetary migration.
\end{abstract}


\section{Introduction}
The number of discovered extrasolar planets is increasing rapidly. 
At the time when this letter was submitted, the number of planets has reached 784\footnote{\url{http://exoplanet.hanno-rein.de}}. 
Of those, 275 \rev{(35\%)} are in multi-planetary systems with two or more planets orbiting the same star. 
These systems are of particular interest to theorists as they can provide valuable information about their formation history.

The existence of mean motion resonances (MMRs) has been confirmed in multiple systems.
The most studied planetary system in a MMR is Gliese~876.
It consists of three gas giants which are locked in a tight 1:2:4 Laplace resonance \rev{\citep{Rivera2010,MarcyButler01}}.
A large number of studies \citep[e.g.][]{LeePeale01,LeePeale2002,SnellgrovePapaloizouNelson01,NelsonPapaloizou2002,beamic2003,veras2007} suggest that migration driven by a variety of mechanisms has played an important role in shaping this system.

This is not surprising as planet migration is a natural outcome of the interaction of a planet with the proto-planetary disk that it forms in \citep{GoldreichTremaine1980}.
Both $N$-body and hydrodynamical models can easily reproduce the observed period ratio, the eccentricities and the libration pattern even though the precise speed at which migration occurs is still up for debate.
Furthermore, it is possible to place limits on the strength of additional stochastic forces which might be present in a turbulent proto-planetary disk \rev{\citep[e.g.][]{ReinPapaloizou2009}}.

Most of these planets have been discovered with the radial velocity method.
This method is biased towards finding heavy planets on close-in orbits. 
Since 2009, the Kepler spacecraft is monitoring over one hundred thousand stars. 
Kepler has discovered thousands of planet candidates which now await confirmation \citep{Batalha2012}.
These planets have on average a much smaller mass than those discovered by radial velocity.
Kepler therefore opens a new window to test planet formation scenarios.

In this letter, we apply a model of smooth migration (i.e. Type~I or Type~II depending on disk and planet properties) as well as stochastic migration forces to each candidate system with multiple planets.
We show that even a small amount of \rev{smooth} migration produces a distribution of period ratio that is inconsistent with observations.
At each resonance, planets pile up, leading to distinct features in the cumulative distribution function of period ratios.

When stochastic migration forces are added, these features are smeared out. 
By adding just the right amount, one can retain some of the features, leading to a period ratio distribution similar to the observed one.
Our simulations show a pile-up just outside of the exact commensurability. 
This is also seen in the Kepler data.
However, it is less apparent in our simulations.
This could either be resolved by fine-tuning several parameters or by adding additional physics such as tidal damping or gap edges.

\section{Setup}
We use the \reb code \citep{ReinLiu2012} to simulate the orbital evolution of planetary systems.
We add dissipative forces to the equations of motion which mimic the interactions of a planet with a proto-planetary disk \citep{LeePeale01}.
This setup allows us to choose two dimensional parameters for each planet, the migration timescale $\tau_a$ and the eccentricity damping time-scale $\tau_e$. 
\rev{Most Kepler planets are not massive enough to open a gap in a minimum mass solar nebula \citep{Weidenschilling1977,Crida2006,Crida2009} and migrate in the Type I regime \citep{Ward1986}.
We therefore choose $\tau_a$ to be a typical value of the Type I migration rate for these planets, $10^3-10^4$~years.
Note that many effects such as a non-isothermal disk may change the Type I migration rate \citep{Paardekooper2010}.  }
We set the eccentricity damping time-scale to be 10 times shorter than the migration timescale in all our simulations,~$\tau_e=\tau_a/10$.
This choice has been adopted be many authors in the past \rev{and has been justified by hydrodynamical simulations \citep{KleyPeitz04}.
Tests have shown that the results do not strongly depend on this precise value.}
We use a 15-th order RADAU integrator with a time-step set to $10^{-3}$ times the innermost planet's period. 
\rev{Further} tests have shown the the results do not depend on the choice of integrator or time-step.

To study the effects of migration, we setup systems that closely resemble those observed by Kepler.
In fact, we initialize the entire set of 364 multi-planetary systems and take the stellar mass, the planet periods and the planet radii directly from the published KOI tables. 
From transit observations only the planet radius is known.
We follow \cite{Fabrycky2012} and assume a simple mass-radius power-law to get a reasonably estimate of the planet mass: $\nicefrac{m_p}{M_\oplus} = \left(\nicefrac{r_p}{R_\oplus}\right)^{2.06}.$
Each planet is initialized on a circular orbit.
All systems are coplanar. 
This seems a reasonable assumptions given that most Kepler planets are expected to be highly aligned \cite[e.g.][]{TremaineDong2012,Sanchis2012}. 
Almost all systems are stable for at least $10^4$ years when initialized this way and we ignore those few ($<1\%$) that are unstable.
We also do not take into account that several KOI objects have already been confirmed and now have improved orbital fits.
We are confident that this procedure sets up systems that are indeed similar to the real planetary systems, at least in an average sense.
We have furthermore tested that a perturbation of the initial orbital parameters (such as starting with a more hierarchical period ratio distribution) does not change the outcome of our simulations.

\section{Smooth Planetary Migration}
\begin{figure}
\centering \resizebox{0.99\columnwidth}{!}{\includegraphics{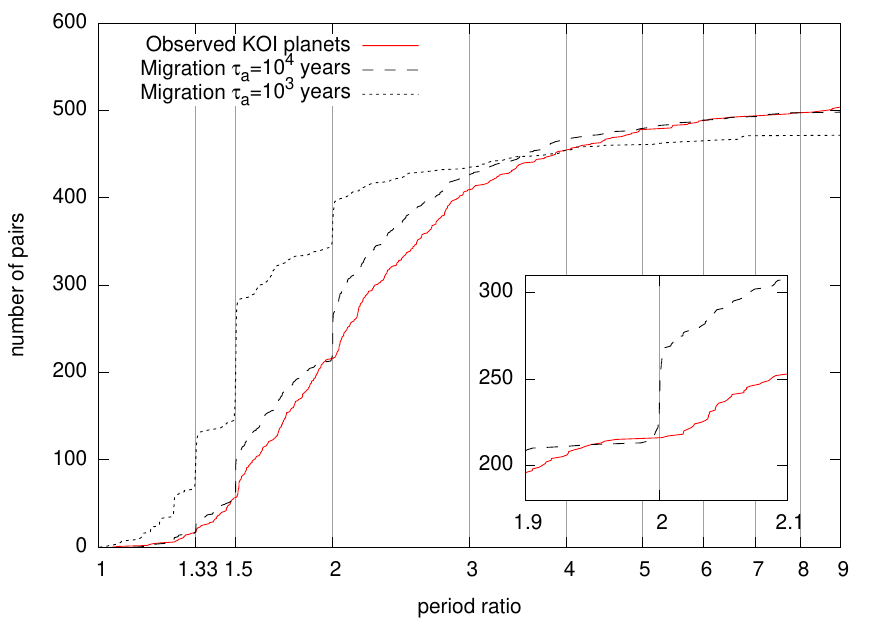}}
\caption{Cumulative distribution of the period ratios in KOI systems with multiple planets; solid (red) line: observed period ratios, long dashed line: period ratios after migration with $\tau_a=10^4$~years, short dashed line: period ratios after migration with $\tau_a=10^3$years. \label{fig:cdf}}
\end{figure}
We add smooth, dissipative migration forces to the equations of motions for the outermost planet in each system. 
All other planets only feel the gravitational forces from the star and the other planets. 
This setup naturally leads to convergent migration. 
As the outer planet moves in, it captures the other planets in resonance. 
Which resonance is chosen depends on the initial position, the migration speed and the planet masses \rev{\citep{Mustill2011b}}.
We stop the integrations after $10^4$ periods of the outer planet at it's initial location.
By that time it has moved in significantly.
We tested removing the migration forces after a different amount of time but did not see any qualitative difference.

In Figure~\ref{fig:cdf} we show the cumulative distribution of period ratios of neighboring planets in all observed KOI systems as a solid (red) curve.
Note that there are surprisingly few features near integer ratios\footnote{\cite{ReinPayne2012} present a similar plot for radial velocity system.}.
However, by closely inspecting the curve near 3:2 and 2:1 period ratios, one can see a deficit of planets exactly in the commensurability and a slight pile-up just outside.

The dashed curves show the final period ratios in our integrations. 
We plot the results for two different migration rates, $\tau_a=10^4$~years and $\tau_a=10^3$~years.
There are now very clear and sharp features that can be associated with resonances. 
Most planets end up in the 3:2 resonance followed by the 2:1 and 4:3~resonance.
This is clearly not consistent with the observed distribution.
By closely inspecting each of these locations, one can see a tiny asymmetry favoring larger period ratios over the exact commensurability. 
However, this is nowhere near the very apparent deficit of planets in the observed period ratio distribution.

\section{Stochastic Planetary Migration}\label{sec:stochastic}
\begin{figure}
\centering \resizebox{0.99\columnwidth}{!}{\includegraphics{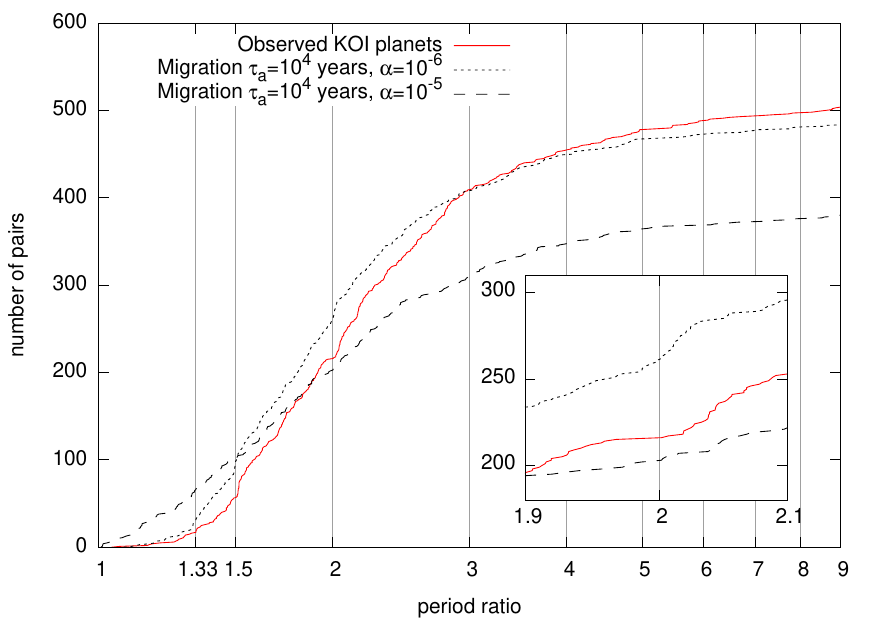}}
\caption{Cumulative distribution of the period ratios in KOI systems with multiple planets; solid (red) line: observed period ratios, long dashed line: period ratios after migration with stochastic forces with $\alpha=10^{-6}$, short dashed line: period ratios after migration with stochastic forces with $\alpha=10^{-5}$. \label{fig:cdf_stochastic}}
\end{figure}
It is very likely that some kind of stochastic force was acting on planets at least during some parts of their past. 
These forces could result form MRI turbulence in the proto-planetary disk \rev{\citep[e.g.][]{ReinPapaloizou2009, Gressel2011}}.
\rev{Another possibility is the gravitational interaction with a remnant planetesimal disk.
This scenario also leads to migration containing both a smooth and stochastic component \citep{Ormel2012}.}
Even the interaction with other planets can be described as a random walk in certain cases \rev{\citep[see e.g.][]{Zhou2007,WuLithwick2011}}. 
Each of these processes is not completely explored and it is hard to estimate the precise amplitude (or the diffusion coefficient) of the stochastic forces in each of those scenarios.
We therefore parameterize the forces in this study.

Stochastic forces are modeled following the procedure described in \rev{\cite{ReinPapaloizou2009}}. 
The radial and azimuthal component of the stochastic force are modeled independently as a Markov process.
The forces have an exponentially decaying autocorrelation function with a finite correlation time.
This mimics the forces of a turbulent accretion disk. 
We set the correlation time to be half of the orbital period of the outer planet. 
The strength of the forces are measured relative to the gravitational force from the central star by the dimensionless parameter $\alpha$. 
\rev{\cite{ReinPapaloizou2009} estimate the value for $\alpha$ to be $\sim5\cdot10^{-6}$ for small mass planets that are embedded in a fully MRI turbulent disk (see their Section 3.1).
It is important to point out that this value is not well constrained.}
In our model each planet is forced stochastically, whereas only the outer planet feels a net inward (Type I/II) migration force as described in the previous section.

In Figure~\ref{fig:cdf_stochastic} we show again the cumulative distribution of period ratios.
The solid (red) line is the observed distribution. 
The dashed lines are from simulations where both smooth and stochastic migration forces are turned on. 
The long and short dashed lines correspond to $\alpha=10^{-6}$ and $\alpha=10^{-5}$ respectively.

One can see that when the stochastic forces are sufficiently strong, $\alpha=10^{-5}$, every resonant feature in the distribution is lost.
Furthermore, several planetary systems become unstable and several planets get ejected (note that the total number of pairs is smaller).
\rev{The number of ejected planets cannot be easily inferred from the observed distribution of (not ejected) planets.}
\rev{Additional simulation with no net migration ($\tau_a=\infty$, not plotted) show the same behaviour.} 

For a small level of stochastic forcing, $\alpha=10^{-7}$, most resonant feature remain. 
This is not shown on the plot as it looks almost identical to the long dashed line in Figure~\ref{fig:cdf}.
This is once again not consistent with the observed distribution.

However, for a more moderate level of stochastic forcing, $\alpha=10^{-6}$, some of the resonant features remain.
Especially the features near strong resonances such as 4:3 and 2:1 can still be seen.
This looks surprisingly similar to the observed distribution.
There is a pile-up of planets just outside the exact commensurability. 
However, one should note that this model still produces slightly too many planets that are too close to the exact commensurability.
But the tendency of planets to have slightly larger period ratios is reproduced correctly.

Finally, note that the overall net migration of planets moves the distribution to the left.
This should not be misunderstood as a prediction but rather illustrates the fact that we start the planetary systems at their current location and then migrate them closer together.
Eventually, one should use an initial distribution that is motivated by models of planet growth and then integrated forward in time until it resembles the observed distribution. 
But this kind of population synthesis introduces many new parameters and we therefore decided not to go down this route.

\section{Eccentricity distribution}
\begin{figure}
\centering \resizebox{0.99\columnwidth}{!}{\includegraphics{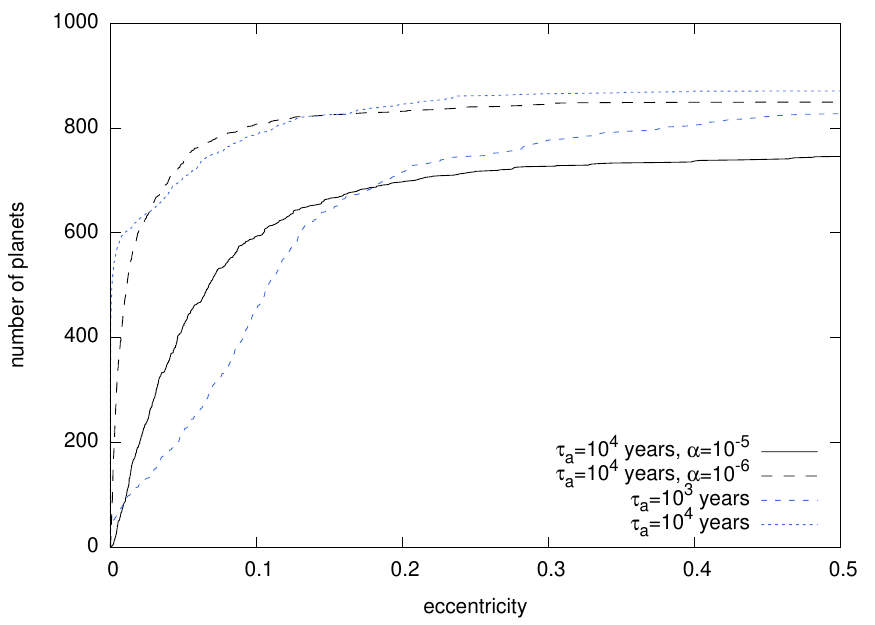}}
\caption{Cumulative distribution of the final eccentricities in runs with $\alpha=0, 10^{-6}$ and~$10^{-5}$, and $\tau_a=10^{3}$~and~$10^{4}$. \label{fig:eccentricity}}
\end{figure}
When planets migrate and capture into resonances, their eccentricities rise. 
Eventually the excitation from the convergent migration and the eccentricity damping from the disk reach an equilibrium \citep{Papaloizou2003}.
Stochastic forces also cause the planets eccentricities to grow \rev{\citep{ReinPapaloizou2009}}. 

If the combination of the smooth and stochastic migration scenarios, as presented in Section~\ref{sec:stochastic}, is indeed responsible for the observed period ratio distribution, then we can make a prediction for the eccentricity distribution. 
The mean eccentricity in our simulation with parameters $\tau_a=10^4$~years and $\alpha=10^{-6}$ is $\langle e\rangle\sim0.01$.
In the simulation with parameters $\tau_a=10^4$~years and $\alpha=10^{-5}$ we have $\langle e\rangle\sim0.05$.

In Figure~\ref{fig:eccentricity} we plot the final cumulative eccentricity distribution in our simulations.
The run with a short migration time-scale, $\tau_a=10^3$~years, produces high eccentricity planets.
Planets are captured into resonance earlier and eccentricities grow faster.
The simulations without stochastic forcing, illustrated by a short and medium dashed curve (blue), show signs of a bimodal distribution.
This is because planets that do (do not) capture in resonance have a high (low) eccentricity.
Finally, note that the run with $\alpha=10^{-5}$ also leads to high eccentricities.
This is due to the random excitation and not due to resonances.

Unfortunately eccentricities are not known for most of the Kepler candidates. 
Only RV follow-up observations or transit-timing variations allow a measurement of the eccentricities. 
However, the eccentricity distribution of multi-planetary systems can be estimated statistically. 
\cite{TremaineDong2012} and \cite{Moorhead2011} \rev{report values of $e$ between $\sim0.1-0.25$.}

It is unfortunate that the simulation which gives the best fit to the period ratio distribution is not the best solution in terms of the mean eccentricity. 
However, it is worth pointing out that observed eccentricities have the tendency to be overestimated and the actual eccentricities might be lower.
\rev{This is a well known effect in radial velocity observations \citep{Zakamska2011} and most likely also present in the Kepler sample.
For example noise in the data or non-transiting planets creating additional transit timing variation might lead to an over-estimate of the mean eccentricity.
The underlying reason for this is that the eccentricity is a positive definite quantity.
}
In addition to that, there are multiple possible \rev{physical} solutions to this discrepancy. 
First, one can fine-tune the eccentricity damping and migration timescales.
The ratio was fixed in our simulations. 
\rev{Further experiments have shown that this ratio indeed has an influence on the final eccentricity distribution, but it is not possible to push the mean eccentricity all the way up to $\sim0.2$.}
Second, the length of our integrations was rather short. 
Longer integration times lead to higher eccentricities due to secular interactions and the still active stochastic force.
Third, massive, non-transiting planets in the systems might also alter the eccentricity distribution. 
All of these effects tend to increase the final eccentricity\rev{, thus going in the right direction}.

A complete model that takes all these effects into account will be able to constrain both smooth and stochastic migration forces.
This is interesting because Type I migration timescales are hard to predict as they depend strongly on local disc properties \citep[see e.g.][]{Paardekooper2010}.
A similar problem exists with stochastic forces.
Dead zones might lower the amplitudes in some regions of the disk.
However, a precise modeling of these effects goes beyond the scope of this letter.

\section{Discussion}\label{sec:discussion}
\begin{figure}
\centering \resizebox{0.99\columnwidth}{!}{\includegraphics{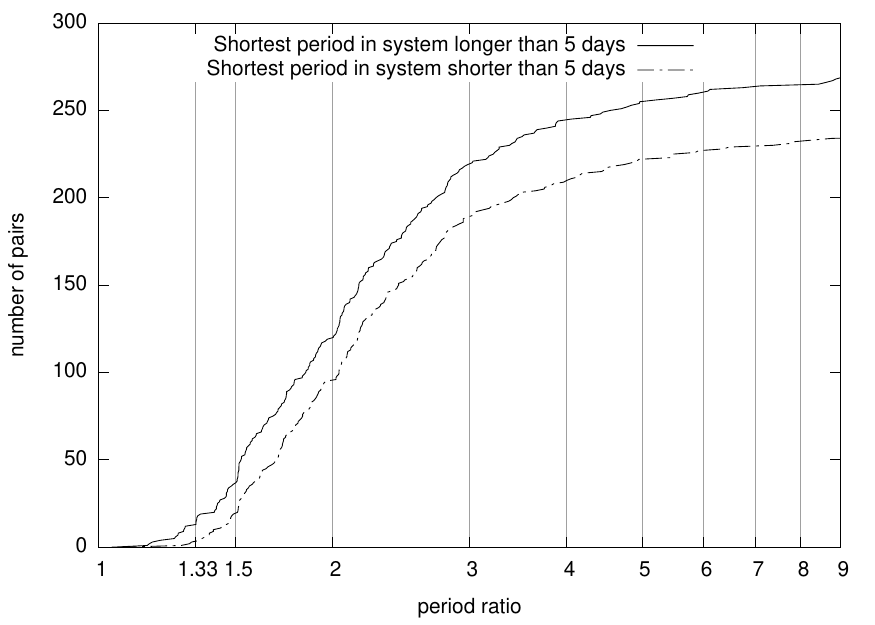}}
\caption{\rev{Cumulative distribution of the period ratios in KOI systems divided into two bins: planetary systems where the period of the innermost planet is shorter/longer than 5 days. \label{fig:cdf_innerperiod}}}
\end{figure}

Our results show that neither smooth nor stochastic planetary migration alone can reproduce the observed period ratio distribution of multi-planetary systems in the Kepler sample. 

However, a combination of those two effects can create a period ratio distribution which is similar to the observed one.
If this scenario is true, then we can use the eccentricity distribution of Kepler planets to constrain the relative strength of stochastic and smooth migration forces. 

Other ideas have been brought up to explain the observed period ratios.
\cite{TerquemPapaloizou2007} show that migration of multiple planets in a disk with an inner edge, together with orbital circularization causes strict commensurability to be lost.
\rev{Similar scenarios involving tidal interactions have been studied by \cite{Delisle2012,BatyginMorbidelli2012,LithwickWu2012}.} 
There is one important difference to the migration scenario presented here.
The inclusion of tides leads a strong radial dependence of the effect.
Such a dependency is currently not observed in the Kepler data set.
\rev{This can be verified in Figure~\ref{fig:cdf_innerperiod} where the cumulative distribution is divided into two bins with almost equal number of planets. 
The two bins contain systems where the innermost planet's period is shorter/longer than 5 days.
If tides were an important factor in shaping this distribution, one would expect a bigger effect for the bin that contains close-in planets.
However, both distributions are identical.
} 

Although this letter focused on the migration scenario, we are interested in seeing other ideas being tested in a framework similar to the one presented in this letter.
We therefore decided to make the initial conditions, the integrator and all plotting routines freely available. 
They can be downloaded as a tar-file from \url{https://github.com/hannorein/rebound/tree/resonancelocation}. 
The relevant files for this project are located in the directory \texttt{problems/resonancelocation}.
We hope that this enables the community to further investigate these extremely interesting planetary systems and come up with new and maybe even better ideas.

\section*{Acknowledgments}
\rev{This letter benefited greatly from comments by the reviewer, Alexander Mustill.}
Hanno Rein was supported by the Institute for Advanced Study and the NFS grant AST-0807444.
Hanno Rein would like to thank \rev{Willy Kley,} Dave Spiegel, Subo Dong and Scott Tremaine for helpful comments.
Dave Spiegel and Subo Dong helped to create a consistent set of initial conditions including the stellar mass and the planet radii for each KOI object.

\bibliographystyle{mn2e}
\bibliography{full}

\end{document}